\documentclass[aps,prd,superscriptaddress,floatfix]{revtex4-1}
\usepackage{color}
\usepackage{amssymb}
\usepackage[normalem]{ulem}
\usepackage{graphicx,color}
\usepackage{amsmath}
\usepackage{subfigure}
\usepackage[normalem]{ulem}
\usepackage{booktabs}
\usepackage{xcolor}
\usepackage{epsfig,graphics,color,graphicx,amsmath}

\colorlet{darkgreen}{green!50!black}
\colorlet{brightyellow}{yellow!75!red}
\colorlet{orange}{red!50!yellow}
\colorlet{darkblue}{blue!60!black}
\colorlet{darkred}{red!80!black}

\usepackage{soul}

\usepackage{mathtools}
\usepackage{mathrsfs}
\usepackage{bm}
\usepackage{relsize}
\usepackage{indentfirst}

\usepackage{xcolor}

\usepackage{amssymb,amsmath,multirow,epsfig,graphicx,color}

\usepackage{epsfig}
\usepackage{graphicx,amsmath}
\def\be{\begin{eqnarray} &&}

	\def\ee{\end{eqnarray}}

\newcommand\ba{\begin{eqnarray}}
	\newcommand\ea{\end{eqnarray}}

\newcommand{\bas}{\begin{eqnarray*}}
	\newcommand{\eas}{\end{eqnarray*}}

\newcommand{\bno}{\begin{eqnarray*}}
	\newcommand{\eno}{\end{eqnarray*}}

\def\sl
\usepackage{hyperref}
\usepackage{url}
\usepackage{hyperref}
\hypersetup{
	colorlinks=true,  
	linkcolor=blue,   
	filecolor=magenta,      
	urlcolor=blue,    
	citecolor=blue,   
} 
\begin{document}
	\vspace{-12ex}    
	\begin{flushright} 	
		{\normalsize \bf \hspace{50ex}}
	\end{flushright}
	\vspace{11ex}
	\title{Modified Uncertainty Principle with Cosmological Constant:\\ More Insights on Dark Energy and Chandrasekhar Limit
	}
	\author{S. Ahmadi}
	\email{samira666ahmadi@gmail.com}
	\affiliation{Department of Physics, Ayatollah Amoli Branch, Islamic Azad University, Amol, Iran}
	\author{E. Yusofi}
	\email{e.yusofi@ipm.ir}
	\affiliation{Department of Physics, Ayatollah Amoli Branch, Islamic Azad University, Amol, Iran}
	\affiliation{School of Astronomy, Institute for Research in Fundamental Sciences(IPM), P. O. Box 19395-5531,Tehran, Iran}
	\affiliation{Innovation and Management Research Center, Ayatollah Amoli Branch, Islamic Azad University, Amol, Mazandaran, Iran}
	
	\author{M. A. Ramzanpour}
	\email{m.ramzanpour212@gmail.com (Corresponding Author)}
	\affiliation{Department of Physics, Ayatollah Amoli Branch, Islamic Azad University, Amol, Iran}
	\affiliation{Innovation and Management Research Center, Ayatollah Amoli Branch, Islamic Azad University, Amol, Mazandaran, Iran}
	
	\date{\today}
	\begin{abstract}
Numerous studies have shown that generalized uncertainty
principle (GUP) removes the Chandrasekhar limit, which can be restored using a negative GUP parameter. This study indicates that observational phantom dark energy also requires an extended uncertainty
principle (EUP) parameter with the opposite sign. Altering the signs of the GUP and EUP parameters without a physical rationale is questionable. We demonstrate that incorporating cosmological constant in a modified uncertainty principle (MUP) can address the sign change in GUP and EUP within a unified framework. The main advantage of MUP is that the sign change of the cosmological constant is acceptable and geometrically meaningful. To achieve this, we first derive a modified equation of state from the MUP framework and second test it with observational data for dark matter and dark energy.  Importantly, the proposed MUP parameter, which is proportional to the cosmological constant with positive and negative signs, aligns with dark energy observations and restores the Chandrasekhar limit for stars. Finally, we will show that the Chandrasekhar mass limit provides an upper bound of \(\leq 10^{-32}{\rm m^{-2}}\) for the cosmological constant, consistent with the observational value of \(\Lambda_{\rm obs}=10^{-52}{\rm m^{-2}}\).
		
	\noindent \hspace{0.35cm} 
	\noindent \hspace{0.35cm} \\
	
	\end{abstract}

	\maketitle
\section{Introduction}

The idea of using a modified form of the Heisenberg Uncertainty Principle (HUP) has emerged as a fundamental concept in modern theoretical physics, addressing the upper and lower limits in position and momentum measurements. These modifications have been applied across various fields, including quantum gravity, special relativity, and cosmology. This development has led to the formulation of the generalized uncertainty principle (GUP), with several variants being introduced~\cite{Maggiore:1993rv, Kempf:1993bq, Hossenfelder:2003jz, Bambi:2007ty, Ali:2009zq}

One of the earliest versions is the GUP, known for defining a minimum length uncertainty in position measurements, which has been extensively investigated in the context of quantum mechanics~\cite{Tawfik:2014zca}. Additionally, other GUP frameworks, which consider both minimum length and maximum momentum, have also been proposed ~\cite{Pedram:2011gw, Pedram:2012my}. This framework studies the implications of GUP on various physical phenomena, such as black hole physics and cosmology, and has been shown to help avoid the singularity associated with the Big Bang~\cite{Scardigli:1999jh,Perivolaropoulos:2017rgq,Mohammadi:2015upa,Nozari:2006au}. Furthermore, several articles have analyzed the thermodynamic properties of systems, including perfect gases and photons, within this framework~\cite{Fityo:2008zz,FarhangMatin:2015lja,Abbasiyan-Motlaq:2014kqa}.

Perivolarpoulos~\cite{Perivolaropoulos:2017rgq} introduced an Extended Uncertainty Principle (EUP), which includes a maximum length $l_{\rm max}$. He highlights important motivations for this type of GUP, especially in the case of infrared cutoff. In fact, the non-local characteristics of quantum mechanics lead to observable effects at large scales. Moreover, the maximal measurable length appears naturally in the context of particle horizons in cosmology or non-trivial cosmological topology. For comprehensive details, please refer to \cite{Perivolaropoulos:2017rgq} and its references. A key question is how a maximum cosmic horizon length impacts the complexities of small-scale physics related to the uncertainty principle. Many sources suggest framing the EUP as a local curvature correction~\cite{CostaFilho:2016wvf,Golovnev:2003xy,Schurmann:2018yuz,Pachol:2024hiz}, which offers a clearer and more intuitive understanding of its role in local physics, rather than depending on the abstract notion of maximum length. To further explore the implications of introducing a maximum length, Bensalem and Bouaziz~\cite{BENSALEM2019583} examined the effects of the maximum length uncertainty principle on the thermodynamic properties of an ideal gas. Especially, it has been shown a modified equation of state, similar to that of real gases, emerges in the scope of EUP formalism.

Recently, in ~\cite{Ahmadi:2024ono}, we investigated the cosmological constant problem and the EUP in a unified framework in a void-dominated scenario for cosmology~\cite{Yusofi:2022hgg, Mohammadi:2023idz, MOSHAFI2024101524}. The analysis indicated there is a substantial discrepancy of approximately \(\mathcal{O}(+122)\) in the cosmological constant values based on varying bubble (void) radii in the cosmos. Also, there is a similar discrepancy in the parameters of EUP and GUP when incorporating both minimum and maximum lengths. Consequently, a novel form of the modified uncertainty principle (MUP) is proposed that incorporates a scale-dependent non-zero cosmological constant~\cite{Ahmadi:2024ono}. In the present study, we will obtain a corrected equation of state and entropy for the real gas by combining the findings of ~\cite{BENSALEM2019583} and \cite{Ahmadi:2024ono}. We examine how our modifications relate to the observed values of the dark energy equation of state and the Chandrasekhar limit for white dwarfs ~\cite{Ong:2018zqn,Ong:2018nzk,Parsa:2024pfn}.

In Section \ref{Section II}, we review the modified uncertainty principle that includes the cosmological constant, considering both maximum and minimum lengths. Section \ref{Section III} presents a method to adjust the properties of an ideal cosmic gas to match its actual behavior, incorporating first-order corrections using the partition function related to maximum length and the cosmological constant. In Section \ref{Section IV}, we analyze modified equations of state and entropy in relation to observations of the dark energy equation of state. Our study concludes with a discussion on restoring the Chandrasekhar limit and establishing an upper bound for the cosmological constant in such high-energy regions within the MUP framework.
\section{Modified Uncertainty Principle with Cosmological Constant}
	\label{Section II}
This section reviews MUP with a scale-dependent cosmological constant, first introduced in~\cite{Ahmadi:2024ono}. MUP, derived from the commutation relation of position and momentum operators, is expressed as follows, 

\begin{equation}
	[X,P]=\frac{\hbar}{2}(1+\alpha X^2+\beta P^2).
\end{equation}

Different representations for the position ($X$) and momentum ($P$) operators can be selected. For example, we can write,  

\begin{equation}  
	X = x \quad \text{and} \quad P = p \left( 1 + \alpha x^2 + \beta p^2 \right) \tag{2}  
\end{equation}  

Under this definition, the corrected uncertainty principle yields explicit maximum and minimum uncertainties for position and momentum, given by,  

\begin{equation}  
	\label{GEUP}
	\Delta X \Delta P \geq \frac{\hbar}{2}  \frac{1}{1- \beta \Delta P^2} \frac{1}{1- \alpha \Delta X^2}~. 
\end{equation}  

Here, $\Delta X$ and $\Delta P$ denote the uncertainties in position and momentum, while the parameters $\alpha$ and $\beta$ are related to the maximum and minimum length uncertainties, respectively.   

Furthermore, by taking the Planck length as the smallest measurable length and the radius of the largest superclusters/supervoids (or horizon radius) as the maximum, we arrive at the following outcome from the relation $
\Delta X\propto \frac{1}{\sqrt{\alpha}}\propto\sqrt{\beta}~$ ~\cite{Ahmadi:2024ono},  

\begin{equation}
	\label{hob19}
	\frac{\alpha_{\rm max}}{\alpha_{\rm min}}\sim\left(\frac{l_{\rm min}}{l_{\rm max}}\right)^{-2}\sim \left(\frac{10^{-35}}{10^{+26}}\right)^{-2} = 10^{+122}.
\end{equation}
This result aligns with the fit for the cosmological constant as $\frac{\Lambda_{\text{max}}}{\Lambda_{\text{min}}} \approx \frac{\Lambda_{\text{planck}}}{\Lambda_{\text{obs}}}=\frac{10^{+70}}{10^{-52}}=10^{+122}$, based on the void-based model \cite{Yusofi:2022hgg}. Given this same behavior of $\Lambda$ and $\alpha$, and similar dimensionality of them as $\alpha \propto \Lambda \sim m^{-2}$, the extended uncertainty principle (\ref{GEUP}) can be reformulated to incorporate the cosmic constant~\cite{Ong:2020tvo,Mignemi:2009ji,Bosso:2023aht} as follow~\cite{Ahmadi:2024ono}, 
\be
\label{GEUPnew}
\Delta X \Delta P \geq \frac{\hbar}{2}  \frac{1}{1- \beta_{0} \Lambda^{-1} \Delta P^2} \frac{1}{1- \alpha_{0} \Lambda \Delta X^2}~.
\ee

In the above relation, $\alpha\equiv\alpha_{0} \Lambda$, and $\beta\equiv\beta_{0} \Lambda^{-1}$ have been used. In relation (\ref{GEUPnew}), it is noted that $\Lambda$ has approximate values, so we need non-zero dimensionless coefficients $\alpha_{0}$ and $\beta_{0}$. This new extension of the uncertainty principle is directly linked to a non-zero cosmological constant, contributing to a more comprehensive understanding of quantum spacetime at both small and large scales in the universe. While our heuristic method can provide valuable insights into these  effects, they should be viewed as a starting point for more accurate solutions.  	
\section{Modification of Equation of State with Cosmological Constant}
\label{Section III}
The canonical ensemble is defined through the parameters \( N \), \( V \), and \( T \). The partition function of a particle of the system is expressed as follows~\cite{BENSALEM2019583},
\be
Q_1 = \sum\limits_{n}\exp(-\beta E_n)
\ee
where \( \beta = \frac{1}{k_B T} \) is the Boltzmann factor, \( k_B \) is Boltzmann's constant, and \( T \) is the thermodynamic temperature and \( E_n \) is the corrected energy spectrum obtained considering the relation $\alpha=\alpha_{0} \Lambda$ as,

\be
\label{En1}
E_n = \epsilon(1 -\frac{\alpha_{0} \Lambda}{3} V^{2/3})^{-2}, \quad (n = 1, 2, 3, \ldots),
\ee

where \( \epsilon \) is the usual energy spectrum. Using the corrected energy spectrum, and canonical partition function to first order we can obtain modified form of entropy up to first order correction as,
\be
\label{Snew2}
S \approx s - \frac{\alpha_{0} \Lambda}{3} V^{2/3} N k_B .
\ee
Also, through the definition of pressure in the canonical ensemble the following modified relation for the equation of state of an ideal gas is obtained as,
\be
\label{Pnew2}
PV = N k_B T \left(1 - 2\alpha_{0} \Lambda V^{2/3} \right)~. 
\ee
This shows a modified equation of state that includes both the effect of maximum length (quantum gravity effect) and cosmological constant (classical gravity effect). Also, we can rewrite (\ref{Pnew2}) as follows,
\be
\label{wgup}
w_{\mathrm{new}} = w_{\mathrm{old}} \left(1 - 2\alpha_{0} \Lambda V^{2/3}\right)~. 
\ee
Therefore, for dark energy and dark matter we can rewrite relation (\ref{wgup}) as follows,
\be
\label{wnewde}
w_{de}^{\mathrm{new}} =  w_{\mathrm{de}} \left(1 - 2\alpha_{0} \Lambda V^{2/3}\right)  
\ee
For dark matter, the value is typically \( w_{\mathrm{dm}} = 0 \), but in some references, it is considered to be very close to zero with the condition \( w_{\mathrm{dm}} \gtrsim 0 \) ~\cite{Kumar:2012gr,Pan:2022qrr}. Therefore, we can rewrite,
\be
\label{wnewdm}
w_{dm}^{\mathrm{new}} = w_{\mathrm{dm}} \left(1 - 2\alpha_{0} \Lambda V^{2/3}\right)  
\ee

\section{Testing the Modified EoS with Observational Data}
\label{Section IV}
The standard cosmological model assumes dark energy has a value of \( w = -1\). The latest data from CMB observations, particularly from the Planck~\cite{Planck:2018jri}, indicate that \( w \) can be more accurately estimated as,
\be
\label{wobs}
w_{\mathrm{Planck}} = -1.03 \pm 0.03 
\ee
Also, a combination of Planck CMB data, baryon acoustic oscillations, type Ia supernovae, and cosmic chronometer data all agree with a final value of EoS parameter as~\cite{Escamilla:2023oce} 
\be
\label{wobs2}
w_{\rm combine} = -1.013^{+0.038}_{-0.043}
\ee
 indicating a tendency towards the phantom phase, i.e. \( (w_{\mathrm{obs}} < -1) \). 

Now, the correction terms must be chosen so that they align with the observational value of the dark energy equation of state. By examining equations (\ref{wnewde}) and (\ref{wnewdm}), we arrive at the following interesting results:
\begin{enumerate}

\item {\textit{Cosmological constant and the sign of EUP parameter}}:
According to the second term on the right side of equation (\ref{wnewde}) and the observational condition \( (w_{\mathrm{obs}} < -1) \), the cosmological constant $\Lambda$ should take a negative sign. This means having a (quasi-) anti-de Sitter (AdS) phase at cosmic scales, which is completely contrary to the observational data indicating a (quasi-) de Sitter (dS) phase. However, if we assume that the sign of the parameter \( \alpha=\alpha_{0}\Lambda \) was incorrectly chosen from the outset, then under this assumption, the cosmological constant in relation (\ref{wnewde}) can be selected with positive sign, which is also consistent with the observational condition. This incorrect choice of sign has been reported in several other cases, including compatibility with the pressure of neutron star degeneracy and the Chandrasekhar limit ~\cite{Ong:2018zqn,Parsa:2024pfn}. Unlike previous methods, which did not allow for a change in the sign of the $\alpha$ within EUP, our proposed model permits both negative and positive signs for the cosmological constant. Therefore, accepting the change of sign $\Lambda \propto \alpha$ in our method, the equation (\ref{wnewde}) should be rewritten  for $w_{\rm de}=-1$ as follows,
\be
w_{de}^{\mathrm{new}} = -1 - 2\alpha_{0} \Lambda V^{2/3} ~.
\ee
Thus, the equation of state for dark energy, influenced by a positive cosmological constant, will yield a value less than -1, consistent with observational data.

\item {\textit{Possible effect of cosmological constant on dark matter pressure}}: According to the second term on the right side of equation (\ref{wnewdm}), that is modified by accepting the change of sign of the parameter in the presence of maximum length as follows,
\be
w_{dm}^{\mathrm{new}} = w_{\mathrm{dm}} + 2w_{\mathrm{dm}}\alpha_{0} \Lambda V^{2/3} 
\ee
Assuming \( \Lambda > 0 \), which corresponds to having dark energy with dS phase and the dominance of the expansion of voids at cosmic scales, the dark matter also experiences additional pressure from that side ( see Fig.2 ~\cite{Yusofi:2022hgg}). Conversely, with \( \Lambda < 0 \), equivalent to having AdS phase in high-dense local regions, the maximum length condition and presence cosmological constant can increase the pressure of dark matter. As we know, beyond the impact of cosmological constant on large scales, it exerts a significant influence on the Local Universe ~\cite{Courtois:2013yfa}. Also, galaxy groups can be characterized by the radius of decoupling from cosmic expansion ~\cite{Brent_Tully_2015}. Recent findings indicate that incorporating the cosmological constant into analyses predicts higher masses for galaxy groups, including the Local Group~\cite{Benisty:2024tlv}. The new study also explores how the nature of dynamical dark energy influences structure formation, particularly concerning the matter power spectrum and the Integrated Sachs-Wolfe effect~\cite{Reyhani:2024cnr}. 
\item {\textit{Increasing of entropy in the cosmic scales}}:
The corrected entropy relation (\ref{Snew2}) is modified with the acceptance of the change of sign of the parameter $\alpha$ as follows,
\be
S_{\mathrm{new}} = S_{\mathrm{old}} + \frac{\alpha_{0} \Lambda}{3} V^{2/3} N k_B~,
\ee
The tendency for entropy to increase is based on the Second Law of Thermodynamics, which asserts that total entropy in an \textit{isolated system} never decreases over time. On a cosmic scale, when \( \Lambda > 0 \), the universe behaves as an isolated system with dark energy, leading to increased entropy in the void-dominated cosmic scale. Conversely, with \( \Lambda < 0 \), the negative correction term causes cluster-dominated regions to act like non-isolated systems, resulting in decreasing entropy during an AdS phase. Buy the way, the cosmic web exhibits a mix of void-dominated and matter-dominated areas. In reality, the cosmic web features a coexistence of void-dominated and matter-dominated regions. However, on large scales, the predominance of supervoids results in \( \Lambda > 0\) and contributes to an increase in the universe's overall entropy. Pandey in ~\cite{10.1093/mnrasl/slx109,10.1093/mnrasl/slz037} demonstrated that the configuration entropy rate is negative in over-dense regions like sheets, filaments, and clusters, while it is positive in under-dense regions like voids and super-voids. Therefore, it was concluded that merger of cosmic voids can mimic the behavior of dark energy.

\end{enumerate}
\section{Restoring of the Chandrasekhar limit and upper bound for cosmological constant}
Many studies have demonstrated that GUP eliminates the Chandrasekhar limit~\cite{Ali:2013ii, Moussa_2015, Rashidi:2015rro, Mathew:2017drw, Ong:2018nzk}. One way to restore this fundamental limit is by utilizing the GUP parameter with an opposite sign. While altering the sign of the GUP parameter without a physical rationale may be questionable, our MUP model~\cite{Ahmadi:2024ono} provides a coherent framework where the GUP and EUP parameters are proportional to the cosmological constant $\Lambda$. In this context, changing the sign of $\Lambda$ becomes both acceptable and meaningful, potentially leading to intriguing implications in both high-energy physics and cosmology~\cite{Ong:2020tvo}. 

In the MUP framework, a small positive cosmological constant is employed at large scales—consistent with dark energy—while a large negative one is utilized at local high-energy regions. For example, our viewpoints about cosmological constant in MUP are consistent with Ong's work~\cite{Ong:2020tvo}, where he heuristically derived the Schwinger effect in anti-de Sitter (AdS) and de Sitter (dS) spaces using EUP. He found that the EUP parameter must be negative in AdS and positive in dS to yield correct known results.  Our approach can recover the Chandrasekhar limit and predict finite sizes for white dwarfs and neutron stars by allowing the cosmological constant $\Lambda (\propto $ EUP or GUP parameter) to accept both negative and positive signs.

In this final section of the paper, we estimate the numerical value of the $\alpha$ parameter based on our MUP model and compare it with observed values. In our model, we define the following proportional relationships, 
\be \alpha \propto \Lambda \sim V_{\rm s}^{-2/3} \sim R_{\rm s}^{-2}.\ee
In the non-relativistic limit, the radius of a white dwarf $R_{\rm s}$ is related to its total mass $M$~\cite{Ong:2018nzk},
\be R_{\rm s} \sim \frac{\hbar^2}{2Gm_{\rm e}^{\frac{8}{3}}M_{\rm s}^{\frac{1}{3}}}. \ee
Here, the white dwarf is modeled as a pure electron star, while $V_{\rm s}$ denotes the volume, and $m_{\rm e}$ is the electron mass. Thus, we derive the following relationship between $\alpha$ and the stellar mass $M_{\rm s}$, \be \alpha \sim \frac{4G^{2}m_{\rm e}^{\frac{16}{3}}M_{\rm s}^{\frac{2}{3}}}{\hbar^{4}}. \ee 
For a star with mass constrained by the Chandrasekhar limit ($M_{\rm Ch} = 1.44 \, M_{\rm \odot} $), we have
 \be M_{\rm s} \leq M_{\rm Ch}~. \ee 
This leads to the expression 
\be
\Lambda \propto \alpha \leq \frac{4G^{2}m_{\rm e}^{\frac{16}{3}}M_{\rm Ch}^{\frac{2}{3}}}{\hbar^{4}} ,
\ee
where \( G=6.67\times 10^{-11} \, {\rm m^{3} \, kg^{-1} \, s^{-2}} \), \( \hbar=6.6\times 10^{-34} \, {\rm J \, s} \), \( M_{\rm Ch} \simeq 2.765 \times 10^{30} \, {\rm kg} \), and \( m_e=9.1 \times 10^{-31} \, {\rm kg} \). Thus, we can estimate the upper limit for the cosmological constant as 
\be
\Lambda \leq 10^{-32} \, {\rm m^{-2}}.
\ee
 We assumed the white dwarf to be a pure electron star, but the estimated magnitude would not change significantly in reality.

The existence of white dwarfs establishes an upper limit for $\Lambda$ at approximately $10^{-32}{\rm m^{-2}}$, which exceeds the observed value of $\Lambda_{obs}\simeq 10^{-52}{\rm m^{-2}}$, yet remains within the observational range. This upper limit reflects a 102-order-of-magnitude improvement over the natural scale predicted by quantum field theory, around $10^{+70}{\rm m^{-2}}$. Nonetheless, it still falls within the extremes of the cosmological constant, suggesting that its magnitude increases as the scale decreases.
\section{Conclusions}
In conclusion, we have thoroughly investigated the implications of our novel proposed modified uncertainty principle, which incorporates a cosmological constant. Our analysis focused on its effects on the dark energy equation of state and entropy, alongside the observational constraints that arise from this framework. Notably, we compared this modified equation of state to the quadratic form derived from the merger processes of cosmic structures such as clusters and voids, delving into their potential interconnections and the resultant physical implications. Considering the cosmological constant instead of the GUP and EUP parameters enables accurate interpretations of observations on both cosmic and astrophysical scales.

Our findings suggested that the current non-zero value of the cosmological constant, which characterizes dark energy as exerting negative pressure on a cosmic scale, manifested as a positive pressure on clusters and nodes at local scales. This observation hinted at a possible source of dark matter, indicating a deeper relationship between phenomena observed at local and cosmic scales. Furthermore, the proposed method not only aligned with existing dark energy observations but also offered predictions regarding the Chandrasekhar limit and star sizes. It yielded an upper limit of \(\leq 10^{-32} \, \text{m}^{-2}\) for the cosmological constant, consistent with the observational value of \(\Lambda \simeq 10^{-52} \, \text{m}^{-2}\). 

In our proposed modified uncertainty principle and equation of state parameters, the $\alpha$ or $\beta$ parameters do not remove compeletely because the cosmological constant never reaches zero. In the real universe, where the cosmological constant is non-zero, accurate results and predictions cannot be derived from a perfect cosmic fluid. Overall, our work contributes to a more nuanced understanding of the interplay between dark energy, the structure of the universe, and the fundamental principles governing quantum mechanics, paving the way for future research that could further elucidate these complex relationships.

\section*{Declaration of Competing Interest}
	The authors declare that they have no known competing financial interests or personal relationships that could have appeared to influence the work reported in this paper.
	\section*{Acknowledgements}
	This work has been supported by the Islamic Azad University, Ayatollah Amoli Branch, Amol, Iran.\\

	\bibliography{Samira_Paper2.bib}

\end{document}